\documentclass{ws-procs975x65}

\begin{document}

\title{ON SELF-GRAVITATING ELEMENTARY SOLUTIONS OF NON-LINEAR ELECTRODYNAMICS}

\author{J. DIAZ-ALONSO, D. RUBIERA-GARCIA}

\address{LUTH, Observatoire de Paris, CNRS, Universit\'e Paris
Diderot. 5 Place Jules Janssen, 92190 Meudon, France and \\ Departamento de Fisica, Universidad de Oviedo. Avda.
Calvo Sotelo 18, E-33007 Oviedo, Spain}

\begin{abstract}

The electrostatic, spherically symmetric solutions of the general class of non-linear abelian gauge models, minimally coupled to gravity, are classified and discussed in terms of the ADM mass and the electromagnetic energy of the associated flat-space solutions.

\end{abstract}

\bodymatter\bigskip

Generalizations of the Reissner-Nordstr\"om (RN) solution of the Einstein-Maxwell field equations through non-linear electrodynamics (NED) coupled to gravitation have been studied for several decades. This is the case of the Born-Infeld (BI) lagrangian \cite{BI}, originally introduced to obtain a classical theory with finite-energy, electrostatic, spherically symmetric (ESS) solutions. In presence of the gravitational field such solutions were analyzed in several papers \cite{EBI}. Similar analysis have been also performed for ESS solutions of some other models \cite{NED}.

However, while these studies are concerned with \emph{particular} cases of NEDs, the main structure of the gravitating ESS solutions of \emph{general} physically admissible NED models can be largely characterized by the vacuum and boundary behaviours of their lagrangian density functions, no matter their explicit forms everywhere. Let

\begin{equation}
I=\int d^{4}x \sqrt{-g} \left(\frac{R}{16\pi G} - \varphi(X,Y) \right) \label{1},
\end{equation}
be the action of our theory, where $\varphi(X,Y)$ is an arbitrary function of the two field invariants ($X=-\frac{1}{2} F_{\mu \nu}F^{\mu \nu}$, $Y=-\frac{1}{2}F_{\mu\nu}F^{*\mu\nu}$), $F_{\mu\nu}=\partial_{\mu}A_{\nu}-\partial_{\nu}A_{\mu}$ being the field strength tensor and $F_{\mu\nu}^*=\varepsilon_{\mu\nu\alpha\beta}F^{\alpha\beta}$ its dual. We constrain these models by several ``admissibility" conditions: i) $\varphi$ must be single-branched and $C^1$-class on its domain of definition of the $X-Y$ plane, which must include the vacuum. ii) For parity invariance, the condition $\varphi(X,Y)=\varphi(X,-Y)$ must hold. iii) The positive definite character of the energy functional for any field configuration requires \cite{dr09a} $\rho \geq \left(\sqrt{X^{2}+Y^{2}} + X\right) \varphi_X + Y\varphi_Y - \varphi(X,Y) \geq 0,$ where $\varphi_X\doteq \partial \varphi/\partial X$ and $\varphi_Y\doteq \partial \varphi/\partial Y$.

As a consequence of the source symmetry $T_0^0=T_1^1$, and the static spherically symmetric metric may be written as $ds^2=g(r)dt^2+g^{-1}(r)dr^2-r^2(d \theta^2 +\sin^2 \theta d\vartheta^2)$. The field equations $\nabla_{\mu}(\varphi_X F^{\mu\nu}+\varphi_Y F^{*\mu\nu})=0$ for these models lead, for ESS fields ($\vec{E}(r)=E(r)\frac{\vec{r}}{r}, \vec{H}=0$), to a first-integral $r^2 E(r) \varphi_X=q$, which forms a compatible system with the Einstein equations $G_{\mu\nu}=8\pi T_{\mu\nu}$. Its solution can be obtained as \cite{dr09b}

\begin{equation}
g(r)=1-\frac{2m(r)}{r} \label{2},
\end{equation}
where $m(r)=M-\varepsilon_{ex}(r,q)$  contains the ADM mass $M$ and $\varepsilon_{ex}(r,q)=4\pi \int_r^{\infty}R^2T_0^0(R,q)dR$, the (flat-space) energy of the ESS field outside of the sphere of radius $r$. For admissible models $\varepsilon_{ex}(r,q)$ can be shown to be a monotonically decreasing and concave function of $r,$ for fixed values of the electric charge $q$ \cite{dr09b}.

Let us now consider the class of admissible NED models supporting flat-space finite-energy ESS solutions \cite{dr09a}. For the total energy of the electromagnetic field in flat space, $\varepsilon(q)=4\pi \int_0^{\infty}r^2T_0^0(r,q) dr=q^{3/2}\varepsilon(q=1),$ to be finite, we must have fields vanishing at infinity as $E(r)\sim \beta/r^p,$ with $p>1$ ($1<p<2$: case \underline{B1}, $p>2$: case \underline{B3}, and $p=2$: case \underline{B2}; which correspond to a slower than, faster than, and Coulombian behaviours, respectively) while at $r=0$ there are two possible behaviours: \underline{A1}, where $E(r)\sim \beta r^p, -1<p<0$ and \underline{A2}, with $E(r)\sim a-br^{\sigma}$ , $\beta$, $a$ and $\sigma$($>0$) being universal constants for a given model, while $b$ is a function of $q$. For these fields the integral defining $\varepsilon_{ex}(r,q)$ is convergent for any $r$. Let us stress that any admissible NED model supporting flat-space finite-energy ESS solutions belongs to one of the B-cases at $r\rightarrow \infty$ and to one of the A-cases as $r\rightarrow0$. Using \emph{only} these data the associated gravitating ESS solutions can be fully classified. Such a classification can be done by looking for the horizons ($g(r_h)=0$) present in each configuration. The general mass-horizon radius relation, obtained from Eq.(\ref{2}) reads

\begin{equation}
M(r_h)-\frac{r_h}{2}=\varepsilon_{ex}(r_h,q) \label{3},
\end{equation}
and thus the horizon radii (if any) of the different gravitating solutions are defined by the cut points of the curves $\varepsilon_{ex}(r,q)$ (fixed $q$) with the beam of straight lines $M-r/2$, given by different values of the ADM mass $M$. In this sense, a tangency cut point ($g'(r)\mid_{r_h} =0$) leads to an extreme black hole (EBH), defined by the condition $8\pi r^2 T_0^0(r,q)=1$, which gives the EBH mass $M_{extr}(q)$ through Eq.(\ref{3}). For completeness let us also consider the case of NEDs leading to (flat-space) divergent-energy solutions, which can be tackled following the same method as in the finite-energy cases. This procedure leads to the following gravitating structures (see Fig.1):

\begin{figure}
\begin{center}
\includegraphics[width=8cm,height=5.0cm]{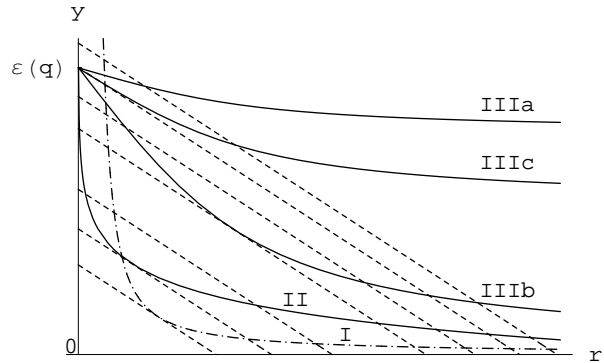}
\caption{\label{fig:epsart} Behaviours of $\varepsilon_{ex}(r,q)$ for the ESS solutions of admissible NED models:
(I) Divergent-energy case; (II) Finite-energy (A-1); (III) Finite-energy (A-2) (IIIa: $16\pi qa<1$, IIIb: $16\pi qa>1$, IIIc: $16\pi qa=1$). The dashed straight lines correspond to different values of $M$ in $M-r_h/2$.}
\end{center}
\end{figure}

\underline{(Flat-space) divergent-energy solutions}: In this case there are three possible structures (see Fig.1, curve I): i) $M=M_{extr}(q)$: EBH, ii) $M<M_{extr}(q)$: Naked singularity (NS), iii) $M>M_{extr}(q)$: Two-horizon (Cauchy and event) BH. Thus the behaviour of any admissible NED model of this class is similar to that of the RN solution of the Einstein-Maxwell equations ($\varphi(X,Y)=\alpha X$, $\alpha$ a constant).

\underline{Finite-energy solutions: Case A-1}: At the center $\varepsilon_{ex}(r,q)\sim \varepsilon-\frac{16\pi q \beta}{(2-p)(1+p)}r^{p+1}$ which implies a negative divergent slope there (see Fig.1, curve II). There are five classes of solutions: i) $M=M_{extr}(q)$: EBH, ii) $M<M_{extr}(q)$: NS, iii) $M_{extr}(q)<M<\varepsilon(q)$: Two-horizon BH, iv) $M=\varepsilon(q)$: Single-horizon BH to which the sequence of solutions of case iii) converges (for $r\neq 0$) when $M-\varepsilon(q)\rightarrow 0^-$, v) $M>\varepsilon(q)$: BH with a single horizon. An example of this family is provided by the Euler-Heisenberg lagrangian, where $\varphi(X,Y)=X/2+\xi(X^2+(7/4)Y^2)$, $\xi>0$ \cite{dr09b}.

\underline{Finite-energy solutions: Case A-2}: At the center $\varepsilon'_{ex}(0,q)=-8\pi q a \gtreqless -1/2$, thus we have three different behaviours for $\varepsilon_{ex}(r,q)$ (see Fig.1, curves III): \textbf{A2a $\leftrightarrow$ curve IIIa)} If $16\pi q a < 1$ there are three cases: i) $M<\varepsilon(q)$: NS, ii) $M>\varepsilon(q)$: Single-horizon BH, iii) $M=\varepsilon(q)$: NS with $g(0)=1-16 \pi qa>0$. \textbf{A2b $\leftrightarrow$ curve IIIb)} If $16\pi q a>1$ we have five cases: i) $M=M_{extr}(q)$: EBH, ii) $M<M_{extr}$: NS, iii) $M_{extr}(q)<M<\varepsilon(q)$: Two-horizon BH, iv) $M>\varepsilon(q)$: Single-horizon BH, v) $M=\varepsilon(q):$ Single-horizon BH with $g(0)=1-16 \pi qa<1$. \textbf{A2c $\leftrightarrow$ curve IIIc)} If $16\pi q a=1$ the charge is fixed and this case is similar to the A2a, expecting for $M=\varepsilon(q)$ where $g(0)=1-16 \pi q a=0$ and we have an ``extreme black point" \cite{dr09b}. The BI model $\varphi(X,Y)=2\beta^2(1-\sqrt{1-\beta^{-2}X-(\sqrt{2}\beta)^{-4}Y^2})$ belongs to this family.

Many other properties of the gravitating ESS solutions of admissible NED models can be established through this procedure. In particular, for the thermodynamic analysis one must take into account that both the zeroth and first laws of BH thermodynamics hold for NEDs \cite{Rasheed}. The Hawking temperature of the ESS solutions is given by $T=\frac{k}{2\pi}; k=\frac{1}{2} g'(r)\mid_{r_{h}}=(\frac{1}{2r_h}-4\pi r_h T_0^0(r_h,q))$ and its analysis leads, aside from ``Schwarzschild-like" and ``RN-like" behaviours, to other cases with very special features. As an example, for $16\pi q a=1$ (A2c case) and $M=\varepsilon(q)$, we find vanishing-$T$($\sigma > 1$), finite-$T$($\sigma = 1$) or divergent-$T$($\sigma < 1$) extreme black points($r_{hextr} = 0$). The analysis of these and other properties of the solutions is currently in progress \cite{dr10}.

\bibliographystyle{ws-procs975x65}

\end{document}